\documentclass[aps,showpacs,twocolumn,pra,titlepage,superscriptaddress,
amssymb,
]{revtex4-1}
\usepackage{amsmath,float}
\usepackage{bm,amsfonts,graphicx,color,amssymb,epstopdf}

\begin{document}

\title{Pulse propagation in a medium optically dressed by a phase-modulated field}

\author{Andrzej Raczy\'nski}
\email{Corresponding author: raczyn@fizyka.umk.pl}
\author{Jaros\l aw Zaremba}
\affiliation{Institute of Physics, Faculty of Physics, Astronomy and Informatics,
Nicolaus Copernicus University, ul. Grudzi\c{a}dzka 5, 87-100 Toru\'n, Poland}
\author{Sylwia Zieli\'nska-Raczy\'nska}
\affiliation{Institute of Mathematics and Physics,
University of Science and Technology, Aleja
Prof. S. Kaliskiego 7, 85-789 Bydgoszcz, Poland}

\begin{abstract}
Pulse propagation is studied in an EIT medium with the control field having a periodically varying phase (chirp).  Based both on numerical calculations and on an approximate approach neglecting absorption and nonadiabatic effects, it is shown that transparency occurs for probe pulses having a sufficiently narrow envelope and an appropriately suited  chirp. For other chirped pulses the evolution of their spectra along a sample is studied. Using a projection method one can divide a pulse into a part for which the medium is transparent and a rest for which it is opaque.
\end{abstract}
\maketitle
\newpage
\section{Introduction}
This work is a contribution to a large family of papers on light propagation in optically dressed media. In the seminal works of Harris and co-workers \cite{harris1,harris2} it was shown that a medium, initially opaque for a resonant probe pulse, might become transparent due to an irradiation by a strong control field coupling the upper state with a long living side state in the so-called $\Lambda$ configuration. This process, called electromagnetically induced transparency (EIT) and described in terms of a transparency window in the medium's electric susceptibility,  has become a basis for numerous studies consisting in its generalizations in various aspects (for reviews see, e.g. Refs. \cite{EIT,lukin,shore1,shore2}). In particular additional active states and coupling fields have been added or the paradigm of a single probe field has been generalized to admit a few coupled probe pulses or control fields. If  one additionally allows the control fields to vary in time by changing their amplitudes or phases one obtains a means to control the propagation of the probe pulse or pulses. A spectacular example is a slowdown of the probe pulse due to an adiabatic decrease of the amplitude of the control field. Switching the latter field off results in mapping the probe pulse into a coherent excitation of the medium, which is called light storage; switching the control field on again leads to a transformation of this excitation back into the probe pulse which is in fact a pulse release with the phase relations being conserved \cite{hau,phillips,tur}.

A qualitatively new situation occurs if the above-mentioned additional level and coupling scheme constitutes a closed interaction contour (loop) or even more loops \cite{kosachiov1,cerb1,cerb2,korsunsky,mahmoudi,eilam1,scully,my13,my14}. In such systems new fields can be generated, the details of the propagation become dependent on the overall phase of the fields and in general steady-state conditions cannot be attained even if all the fields have time-independent envelopes. Within the paradigm of weak probe and given control fields one cannot describe the propagation in terms of a single electric susceptibility and the description of the probe field propagation requires an analysis in terms of coupled channels and a kind of "normal modes". If additionally a strong control field couples the ground state with another state, the populations of atomic states may significantly change which may lead to light generation on the same transition in which the probe field acts.

In all the above-mentioned studies the couplings due to both probe and control fields were resonant or nearly resonant while the manipulations with the phases can lead to off-resonant couplings which can obviously affect EIT. The case of a rapid variation of an optical phase on EIT and the following essential changes in the medium absorption were investigated by Abi-Salloum {\em et al.} \cite{salloum}. This effect was demonstrated experimentally by Sautenkov {\em et al.} \cite{sautenkov2}.

In this paper we focus on the case of a $\Lambda$ configuration in which the  control field has a periodically varying phase (chirp). In other words, using an expansion of the phase factor into the Fourier series, one can say that a probe field propagates in the medium which is dressed by a manifold of control fields with the frequencies differing by the chirp frequency and constituting an infinite number of coupled elementary loops. The resonance for the control field occurs thus during a certain fraction of the modulation period or in other words only for a part of the control field photons. This problem was first studied in the work of Kiffner and Dey \cite{kiffner}. The authors demonstrated that a medium dressed by a chirped control field might in some situations exhibit transparency. They analyzed the problem both numerically and using a simplified model in the case of a small modulation frequency. Their interpretation was based on the concept of a transparency window which oscillated in time. Two particular examples were demonstrated: (1) a continuous probe field which became modulated (chopped) due to the transparency occurring only during a fraction of the control field modulation period and (2) a nonresonant Gaussian pulse which could be transmitted in the situation of a proper timing, i.e. when it arrived just at the time when the window's position suited the pulse's detuning. However, as those authors have noticed,  such an analysis fails for larger modulation frequencies. In particular the modulation of the continuous probe field becomes then more rapid and shallow and disappear when the frequency increases.

In the present paper we consider the  case of a medium dressed by a chirped control field of an arbitrary, in particular large, frequency of the phase modulation. We apply the same approach which was presented in our recent paper on probe pulse propagation in a medium dressed by three control fields constituting a loop \cite{my15}.

In the next section we write the set of the Maxwell-Bloch equations with the probe field treated perturbationally. Next we apply a kind of the Floquet expansion and present the method of solution. In the following section we give approximate analytical solutions, obtained in the spirit of early papers on EIT and light storage, namely assuming an adiabatic and absorptionless pulse propagation. Those analytical formulas for the pulse spectrum together with a discussion of their applicability allow for an interpretation of numerical results presented in the last section. We examine how and in which conditions the idea of EIT can be generalized in the case of chirped control pulses. In particular we find an optimal probe pulse (i.e. such with a suitably chosen chirp) for which transparency occurs and for a pulse which is not optimal we show that its part which is transmitted can be expressed in terms of its projection onto the optimal pulse.

Atomic units are used throughout the paper. In particular the atomic unit of time equals $2.42\times10^{-17}$ s, the atomic unit of length is $0.529\times10^{-10}$ m and the atomic unit of frequency is $6.58\times10^9$ MHz.

\section{General theory}
\begin{figure}[H]
\includegraphics[scale=0.6]{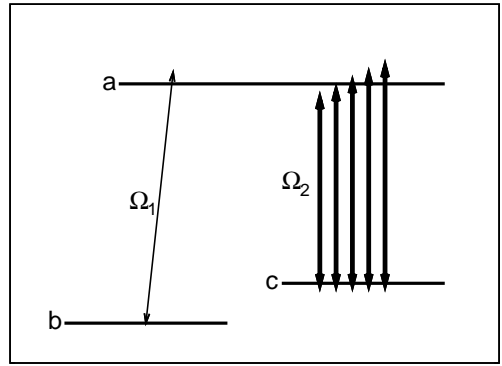}
\caption {\label{fig1} Level and coupling scheme for the considered model. Multiple arrows symbolize the coupling of the states $a$ and $c$ by a periodically chirped pulse, or, alternatively, by a manifold of pulses of frequencies differing by the chirp frequency.}
\end{figure}

The object of our interest is a one-dimensional medium of atoms irradiated by two laser beams in the $\Lambda$ configuration: a probe pulse 1 couples the ground state $b$ with an excited state $a$ while a strong control field 2 couples the state $a$ with a long-living side state $c$. A novel element is that we assume the control field to have a constant amplitude but a chirped, periodically varying phase; this is equivalent to saying the the states $a$ and $c$ are coupled by an infinite number of fields of frequencies differing by a multiple of the chirp frequency (see Fig. \ref{fig1}). One can view this coupling scheme as a ground state being coupled by a weak field $\Omega_1$ with an infinite number of coupled loops.
Because of the probe field being weak the ground state remains fully occupied. For the matrix elements of the atomic density matrix which are nonzero in the first order with respect to the probe field we obtain the following Bloch equations
\begin{eqnarray}
i\dot{\sigma}_{ab}(z,t)&&=(-\delta_1-
i\gamma_{ab})\sigma_{ab}(z,t)\nonumber\\&&-\Omega_1(z,t)-\Omega_2\exp(i\phi(t)) \sigma_{cb}(z,t),\nonumber\\
i\dot{\sigma}_{cb}(z,t)&&=(-\delta_1+\delta_2-i\gamma_{cb})\sigma_{cb}(z,t)\nonumber\\&&-\Omega_2^*\exp(-i\phi(t)) \sigma_{ab}(z,t).
\end{eqnarray}
In Eq.(1) $\sigma$ is the atomic density matrix after the oscillations with the optical frequencies have been transformed off, $z$ is the position of the atom inside the one-dimensional sample, $\Omega_1(z,t)$ and $\Omega_2=const$ are the Rabi frequencies of the probe and control fields, respectively, $\delta_1=E_b+\omega_1-E_a$ and $\delta_2=E_c+\omega_2-E_a$ are the detunings of the two fields, $\gamma_{ab}$ and $\gamma_{cb}$ are phenomenological relaxation rates for the two coherences, $\phi(t)$ is a periodic phase (chirp).
The above Bloch equations are completed by the Maxwell equation for the probe field which takes the following form in the slowly-varying amplitude approximation
\begin{equation}
\big(\frac{\partial}{\partial t}+c \frac{\partial}{\partial z}\big)\Omega_1(z,t)=i \kappa^2 \sigma_{ab}(z,t),
\end{equation}
with $\kappa^2=\frac{{\textsf N}|d_{ab}|^2 \omega_1}{2\epsilon_0\hbar}$, ${\textsf N}$ being the atomic density, $d_{ab}$ - the matrix element of the dipole moment and $\epsilon_0$ the vacuum electric permittivity.
The control pulse with a periodically varying phase can be written as
\begin{eqnarray}
\Omega_2 \exp[i\phi(t)]=\Omega_2\sum_{n=-\infty}^{\infty}c_{n}\exp(in\Delta t),
\end{eqnarray}
where $\Delta$ is the chirp frequency and, to assure the phase $\phi$ being real, the coefficient $c_n$ satisfy the condition $\sum_{n}c_n^{*} c_{n+k}=\delta_{k0}$. In what follows we deal with sinusoidal chirps, i.e. such that $\phi(t)=g\sin\Delta t$; in such a case the expansion coefficients are the Bessel functions: $c_n=J_n(g)$ and the control pulse can be expanded as
\begin{eqnarray}
\Omega_2 \exp[ig\sin\Delta t)]=\Omega_2\sum_{n=-\infty}^{\infty}J_{n}(g)\exp(in\Delta t).
\end{eqnarray}
We stress however that our conclusions will hold for any periodical phase.

We pass to the Fourier picture and treat the functions the arguments of which differ by multiples of $\Delta$ as independent variables; this is in fact the Floquet approach. We do not impose any restrictions on the value of  the modulation frequency $\Delta$.  In the Fourier picture the Bloch-Maxwell equations take the form
\begin{eqnarray}
(\omega+\delta_1&&+i\gamma_{ab})\sigma_{ab}(z,\omega)=-\Omega_1(z,\omega)\nonumber\\
&&-\Omega_2\sum_{n=-\infty}^{\infty}J_n(g) \sigma_{cb}(z,\omega+n\Delta),\nonumber\\
(\omega+\delta_1&&-\delta_2+i\gamma_{cb})\sigma_{cb}(z,\omega)=\nonumber\\
&&-\Omega_2^*\sum_{n=-\infty}^{\infty}J_{-n}(g) \sigma_{ab}(z,\omega+n\Delta),\\
&&(-i\omega+c\frac{\partial}{\partial z} \Omega_1(z,\omega))=i\kappa^2\sigma_{ab}(\omega),\nonumber
\end{eqnarray}
where $\Omega$ is the Fourier variable.
Eliminating $\sigma_{cb}$ leads to the system of equations for $\sigma_{ab}$
\begin{eqnarray}
&&(\omega+s\Delta+\delta_1+i\gamma_{ab})\sigma_{ab}(z,\omega+s\Delta)\\
&&-|\Omega_2|^2\sum_{n,k}\frac{J_n(g)J_{n-k}(g)}{\omega+(s+n)\Delta+\delta_1-\delta_2+i\gamma_{cb}}
\times\nonumber\\
&&\sigma_{ab}(z,\omega+(s+k)\Delta)=
-\Omega_1(z,\omega+s\Delta).\nonumber
\end{eqnarray}
Eq. (6) can be written in the matrix form as
\begin{equation}
\sum_k M_{sk}\sigma_{ab}(z,\omega+(s+k)\Delta)=-\Omega_{1}(z,\omega+s\Delta).
\end{equation}
The dimension of the problem is formally infinite but in practice it is truncated, the convergence in general requires a numerical checkup. In the case of a sinusoidally chirped pulse the truncation can be controlled using the fact that the absolute values of the Bessel functions $J_n(g)$ become small if the number $n$ is significantly larger than the argument  $g$.
The coherence $\sigma_{ab}$ can be calculated as
\begin{equation}
\sigma_{ab}(z,\omega+s\Delta)=-\sum_{k}M^{-1}_{sk} \Omega_1(z,\omega+k\Delta).
\end{equation}
Using the last equation in the propagation equation one can write
\begin{equation}
(-i\omega+c\frac{\partial}{\partial z}) \Omega_1(z,\omega+s\Delta)=i\sum_k N_{sk}\Omega_1(z,\omega+k\Delta),
\end{equation}
where the matrix $N_{sk}=s\Delta\delta_{sk}-\kappa^2 M^{-1}_{sk}$.
The solutions of the last equation can be expressed in terms of the eigenvectors and eigenvalues of the matrix $N$ satisfying
\begin{equation}
N(\omega)U(\omega)=U(\omega)N^d(\omega),
\end{equation}
where the superscript $d$ denotes a diagonal matrix of eigenvalues and $U$ is the matrix of eigenvectors.
The field at position $z$ is thus expressed in terms of the incoming field (i.e. at $z=0$) as
\begin{eqnarray}
&&\Omega_1(z,\omega+j\Delta)=\nonumber\\ \sum_{k,s}
&&U_{jk} \exp[\frac{iz}{c}(\omega+N^{d}_{kk})]
U^{-1}_{ks}\Omega_1(0,\omega+s\Delta).
\end{eqnarray}
This result means that the incoming pulse can be considered as composed of a number of modes each of which propagates with a given susceptibility, proportional to $N^d_{kk}$; the modes at a given position are collected to form a final pulse. This gives account of a sequence of absorption and emission processes of virtual photons of frequency $\omega_2+n\Delta$ for various values of $n$, treated in a nonperturbative way.

\section{Approximate solution}

In  this subsection we present approximate solutions of the Maxwell-Bloch equations. They are obtained by generalizing the approach which led to an interpretation of EIT and explaining light storage \cite{fleischhauer2}. As we will see, this approach will allow one to predict the medium's transparency for a chirped pulse in the case in which the chirp is appropriately chosen while the envelope fits the transparency window corresponding to an unchirped pulse.

In the adiabatic ($\dot{\sigma}_{ab}=0$), relaxationless ($\gamma_{ab}=\gamma_{cb}=0$) and resonant ($\delta_1=\delta_2=0$) conditions the coherence $\sigma_{ab}$ can be calculated from Eqs. (1) as
\begin{equation}
\sigma_{ab}(z,t)=-i\frac{1}{\Omega_2^*\exp(-i\phi(t))}\frac{\partial}{\partial t}\frac{-\Omega_1(z,t)}{\Omega_2\exp(i\phi(t))}.
\end{equation}
The propagation equation takes then the form
\begin{equation}(\frac{\partial}{\partial t}+c\frac{\partial}{\partial z}) \Omega_1(z,t)=\frac{-\kappa^2}{|\Omega_2|^2}
(\frac{\partial \Omega_1(z,t)}{\partial t}-i\dot{\phi}\Omega_1(z,t)).
\end{equation}
The solution of the last equation can be written in the form
\begin{equation}
\Omega_1(z,t)=\tilde{\Omega}_1(z,t)\exp(i\alpha(t)),
\end{equation}
where $\alpha(t)$ satisfies the relation  $\alpha(t)=\phi(t)\sin^2\theta$
with $\tan\theta\equiv\frac{\kappa}{|\Omega_2|}$.
The function $\tilde{\Omega}_1$, having usually but not necessarily the character of an envelope, satisfies then the equation
\begin{equation}
(\frac{\partial}{\partial t}+c \cos^2\theta\frac{\partial}{\partial z}) \tilde{\Omega}_1(z,t)=0,
\end{equation}
the solution of which is
\begin{equation}
\tilde{\Omega}_1(z,t)=\tilde{\Omega}_1(z=0,t-\frac{z}{v_{g}}).
\end{equation}
It is important to keep in mind that this solution is valid only if the spectral profile of the function $\tilde{\Omega}_1$ remains inside the transparency window typical of the case of an unchirped control field, so neglecting absorption processes is justified. The transparency window is the frequency range in which the imaginary part of the susceptibility $\chi(\omega)$ given by
\begin{equation}
\chi(\omega)=-\frac{{\textsf N}|d_{ab}|^2}{\epsilon_0\hbar}
\frac{1}{\omega+\delta_1+i\gamma_{ab}-\frac{|\Omega_2|^2}{\omega+\delta_1-\delta_2+i\gamma_{cb}}}
\end{equation}
is negligible; the width of the transparency window is thus of order of $|\Omega_2|$.

The solution $\tilde{\Omega}_2$ describes a function propagating  with the group velocity $v_g=c\cos^2\theta$ with its shape being preserved.
The solution for the pulse $\Omega_1$ reads
\begin{equation}
\Omega_1(z,t)=\Omega_1(z=0,t-\frac{z}{v_g}) \exp[-i\alpha(t-\frac{z}{v_g})]\exp[i\alpha(t)].
\end{equation}
The latter solution represents a pulse shape $|\Omega_1(z,t)|$ which does not change during propagation. However its spectrum exhibits interesting features.

The pulse shape $\Omega_1$ is a product of shape fixed at the entrance to the sample and traveling with the group velocity and of a complicated phase factor. This is a generalization of the usual case of EIT for which the factor does not appear because $\alpha(t)\equiv 0$. The envelope of the pulse shape in space-time given by $|\Omega_1(z,t)|$ preserves its shape but its spectrum $|\Omega_1(z,\omega)|$ is in general a multipeak structure which evolves along the sample.

For a sinusoidally chirped control field and a probe pulse with $\alpha(t)=g\sin^2\theta \sin\Delta t$, using the expansion analogous to that used in Eq. (4), one obtains for the spectrum of the pulse (18)
\begin{eqnarray}
&&\Omega_1(z,\omega+j\Delta)=\sum_{k,s}J_{k-j}(g\sin^2\theta)\times\\ &&\exp[\frac{iz}{c\cos^2\theta}(\omega+k\Delta)]\nonumber
J_{k-s}(g\sin^2\theta)\Omega_1(0,\omega+s\Delta).
\end{eqnarray}
This result can also be obtained using the formalism presented in Eqs. (5-11). Eq. (12) in the Fourier picture takes the form
\begin{eqnarray}
&&\sigma_{ab}(z,\omega)=\frac{\omega}{|\Omega_2|^2}\Omega_1(z,\omega)+\nonumber\\ &&\frac{g\Delta}{2|\Omega_2|^2}
(\Omega_1(z,\omega+\Delta)+\Omega_1(z,\omega-\Delta)).
\end{eqnarray}
Using the latter coherence $\sigma_{ab}$ in the Fourier transform of the propagation equation (2), after a simple manipulation one obtains the set of equations
\begin{eqnarray}
&&(-i\omega+c\cos^2\theta\frac{\partial}{\partial z})\Omega_1(z,\omega+s\Delta)=is\Delta\Omega_1(z,\omega+s\Delta)+\nonumber\\
&&\frac{g\Delta\sin^2\theta}{2}[\Omega_1(z,\omega+(s-1)\Delta)+\Omega_1(z,\omega+(s+1)\Delta)].\nonumber\\
\end{eqnarray}
Eq. (21) has the same structure as Eq.(9) except that the light velocity in vacuum $c$ has been replaced by the group velocity $c\cos^2\theta$ and the new coupling matrix $N'$ has a simple form
\begin{eqnarray}
&&N'_{ss}=s\Delta,\nonumber\\
&&N'_{s\pm1}=\Delta \frac{g}{2}\sin^2\theta.
\end{eqnarray}
The eigenvalues and eigenvectors of such a matrix are $N'^d_{kk}=k\Delta$ and $U'_{sk}=J_{s-k}(-g\sin^2\theta)$. The matrix $U'^{-1}_{kj}=J_{j-k}(-g\sin^2\theta)$.
Using the latter relations one can reduce the analogue of Eq. (11) to Eq. (19).

In particular let us consider the case of a sinusoidally chirped control field and the probe field entering the sample, built of a Gaussian envelope of time width $\tau$  and a suitably chirped factor
\begin{equation}
\Omega_1(0,t)=\Omega_{10}\exp[-\frac{t^2}{\tau^2}+ig\sin^2\theta \sin\Delta t].
\end{equation}
Then the Fourier transform of Eq. (18) or equivalently Eq. (19) leads to the spectrum
\begin{eqnarray}
\Omega_1(z,\omega)=&&
\Omega_{10}\sum_n J_n(g\sin^2\theta )\exp[i(\omega+n\Delta)\frac{z}{c\cos^2\theta}]\times\nonumber\\&&\sqrt{\pi}\tau
\exp[-\frac{1}{4}(\omega+n\Delta)^2\tau^2)].
\end{eqnarray}
 This pulse has such a particular chirp that it propagates unabsorbed provided its width $\frac{1}{\tau}$ is small compared with the width $|\Omega_2|$ of the transparency  window.

If however the entering probe field is a pure Gaussian, i.e.
\begin{equation}
\Omega_1(0,t)=\Omega_{10}\exp[-\frac{t^2}{\tau^2}],
\end{equation}
the spectrum given by Eq. (18) takes the form
\begin{eqnarray}
&&\Omega_1(z,\omega)=\nonumber\\
&&\Omega_{10}\sum_{n,j} J_n(g\sin^2\theta) J_{n-j}(g\sin^2\theta)
\exp[i(\omega+n\Delta)\frac{z}{c\cos^2\theta}]\nonumber\\ &&\times\sqrt{\pi}\tau
\exp[-\frac{1}{4}(\omega+j\Delta)^2\tau^2)].
\end{eqnarray}
An important property of the approximate spectra given by the two last equations is that they are composed of peaks the width of which is the same as of that corresponding to the spectrum of the incoming pulse.

However, note that the solution given by Eq. (26) has a very limited range of application: if the incoming pulse $\Omega_1(0,t)$ is a pure Gaussian (Eq. (25)), the corresponding shape $\tilde{\Omega}_1$ has a  spectrum which usually does not fit the transparency window and thus the assumption of an absorptionless propagation is not satisfied. In such a  case one can consider the pulse being a sum of a part for which  $\tilde{\Omega}_1$ fits the transparency window (in fact a projection of the considered incoming pulse shape onto the pulse shape given by Eq. (23)) and a rest. The latter will be absorbed during the propagation while the former will propagate unabsorbed. In such a case the peaks of the pulse spectrum, initially relatively wide or even overlapping, will be narrowed until they attain the width allowed by the width of the transparency window typical of an unchirped pulse.

\section{Discussion and illustration}
\begin{figure}[H]
\includegraphics[scale=0.6]{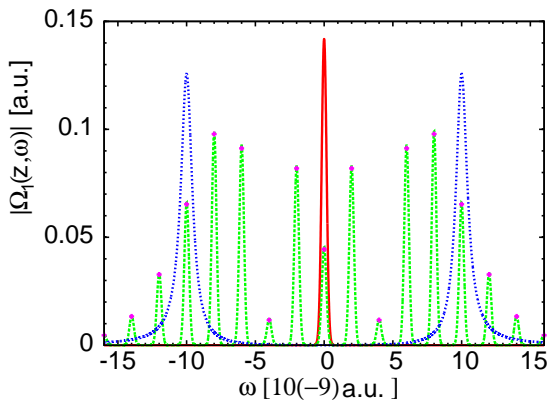}
\caption {\label{fig2} Pulse spectrum for the following data: $\Omega_2=10^{-8}$, $\Omega_{10}=10^{-10}$, $g=5$, $\Delta=2\times10^{-9}$, $\gamma_{ab}=10^{-9}$, $\gamma_{cb}=10^{-14}$, $\delta_1=\delta_2=0$, $\omega_1=10^{-1}$, $\textsf N=2\times10^{-13}$, $d_{ab}=1$,  $z=2\times10^{10}$; the incoming pulse at $z=0$ is a Gaussian of time width $\tau=8\times10^9$ the spectrum of which on the plot has been reduced by the factor of 10 - solid red line; the spectrum of the pulse at the end of the sample $z=2\times10^{10}$, dashed green line; the transparency window corresponding to the unchirped control field, i.e. the imaginary part of the susceptibility multiplied by the factor of 100, dotted blue line; the crosses mark the absolute values of the Bessel functions multiplied by $V\times\Omega_{10}\tau\sqrt{\pi}=0.177\times1.42=0.251$, $V$ being the overlap between the normalized incoming pulses of Figs. \ref{fig1} and \ref{fig3}.}
\end{figure}

Below we present and discuss some examples of the pulse spectrum and shape obtained numerically using Eq. (11); we have adopted the parameters of our model to be of order of those of realistic atomic systems. We start with an example of an incoming Gaussian pulse of spectral width smaller both than the chirp frequency $\Delta$ and the width of the transparency window corresponding to an unchirped control field of the same amplitude. In Fig. \ref{fig2} one can see the outgoing pulse compared with the incoming Gaussian pulse. During the propagation the one-peak structure has gradually been transformed into a multipeak one, with the peaks' heights proportional to the absolute values of the Bessel functions, as expected from Eq. (24) (note that for our data $\sin^2\theta=0.9992\approx 1$). After the pulsed has traveled a sufficiently long way inside the sample its spectrum stabilizes and it remains almost unchanged during a further propagation.

\begin{figure}[H]
\includegraphics[scale=0.6]{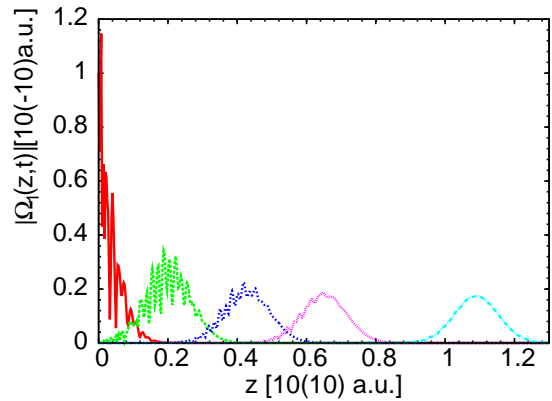}
\caption {\label{fig3} Snapshots of the pulse from Fig. \ref{fig1} taken at $t=0$, solid red line; $t=2\times10^{10}$, dashed green line, $t=4\times 10^{10}$, short-dashed blue line, $t=6\times10^{10}$, dotted violet line; $t=10^{11}$, dash-dotted light blue line.}
\end{figure}

In Fig. \ref {fig3} we show the snapshots of the pulse in space-time. One can see that the pulse is multipeak at the front part of the sample but farther it becomes one-peak, the pulse has a structure of an envelope in space-time multiplied by a phase chirp term which does not influence the absolute value of the pulse.

\begin{figure}[H]
\includegraphics[scale=0.6]{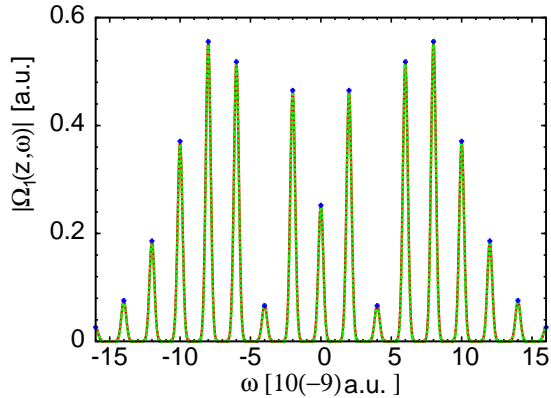}
\caption {\label{fig4}  As in Fig. \ref{fig1} but for an incoming probe pulse ($z=0$) with the chirp depth $g\sin^2\theta$. The spectrum of the incoming pulse, solid red line; that of the outgoing pulse at $z=2\times10^{10}$, dashed green line; the two spectra almost coincide, the crosses mark the absolute values of the Bessel functions multiplied by $\Omega_{10}\tau\sqrt{\pi}=1.42$.}
\end{figure}

In Fig. \ref{fig4} we compare the spectra of the probe pulse at the entrance and at the end of the sample in the case in which the incoming pulse is a suitably chirped one, i.e. it has the form given by Eq. (23) with the width $\tau$, the same chirp frequency $\Delta$ as that of the control pulse and the chirp depth $g\sin^2\theta$ (note again that for our data $\sin^2\theta\approx 1$). Such a pulse propagates with its spectrum as well as the shape in space-time (not shown) almost unchanged from the very beginning of the sample. This agrees with the predictions of Eq. (24) in the case of a spectrally narrow envelope $\tilde{\Omega}_1(0,\omega)$ so that for a given frequency $\omega$ practically only a single term of the sum contributes. Note however that the components of the sum acquire a phase factor which depends on the distance $z$.

The peaks' heights of the outgoing pulse of Figs. \ref{fig2} and \ref{fig4} differ by the value of 0.177 which is the $|J_0(g=5)|$, being simultaneously the overlap between the normalized incoming Gaussian and chirped shapes (for sufficiently spectrally narrow pulses). This means that the Gaussian shape of Fig. \ref{fig2} can be interpreted as being composed of its projection onto the suitably chirped shape (multipeak), proportional to the overlap integral of the normalized pulses, and a rest; the latter becomes transformed and absorbed during the propagation which is an effect taken into account in the general approach leading to Eqs. (5-11) but which is absent from the approximate formalism. The general formula for the overlap of the two normalized chirped pulses having different widths $\tau_j$, depths $g_j$ and chirp frequencies $\Delta_j$ is
\begin{eqnarray}
&&\int_{\infty}^{\infty}dt \exp[\frac{-t^2}{\tau_1^2}-ig_1\sin\Delta_1 t]\exp[\frac{-t^2}{\tau_2^2}+ig_2\sin\Delta_2 t]\times\nonumber\\
&&\sqrt{\frac{2}{\pi\tau_1\tau_2}}=\sum_{n,s}J_n(g_1)J_ s(g_2)\exp[-\frac{(n\Delta_1-s\Delta_2)^2}{4(\frac{1}{\tau_1^2}+\frac{1}{\tau_2^2})}].
\end{eqnarray}

We have also checked that our results hold for a less realistic situation in which $\sin^2\theta$ differs significantly from unity. This required reducing the medium density and extending the sample length by more than two orders of  magnitude. The spectra in such a long sample have been significantly reduced which means that absorption could not be neglected. However, in the cases of the incoming pulses having the chirp depth $g$ and $g\sin^2\theta$ the spectra differed approximately by the value of the overlap of the two pulses (not shown).

\begin{figure}[H]
\includegraphics[scale=0.6]{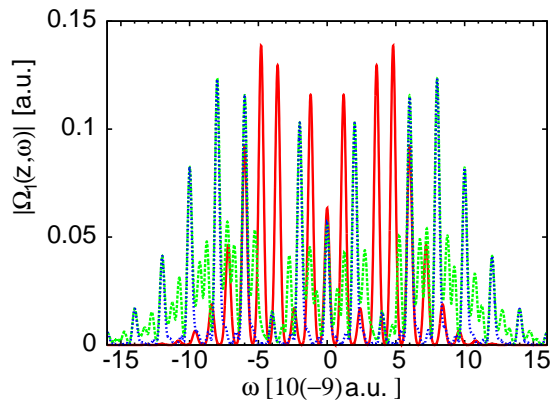}
\caption {\label{fig5}  The pulse spectrum for an incoming pulse of the chirp depth $g'=g\sin^2\theta$ but the chirp frequency $\Delta'=1.2\times10^{-9}$ (different from that of the control field $\delta=2\times10{-9}$), other parameters as in Fig. \ref{fig2}; the spectrum of incoming pulse, solid red line (the values have been reduced 4 times); the spectrum at $z=1.2\times10^{10}$, dashed green line; the spectrum at $z=6\times10^{10}$, short-dashed blue line.
}
\end{figure}

\begin{figure}[H]
\includegraphics[scale=0.6]{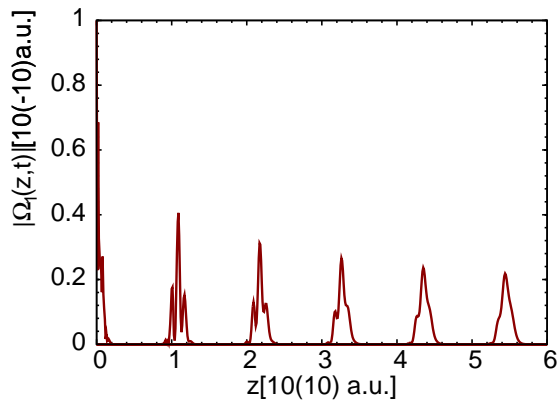}
\caption {\label{fig5b} The snapshots of the pulse from Fig. \ref{fig5} taken at from left to right $t=0,10^{11},2\times10^{11},3\times10^{11},4\times10^{11},5\times10^{11}$.}
\end{figure}

An adjusting of the spectrum of the propagating pulse to the spectrum given by Eq. (24) occurs also in more general cases, i.e. for chirps of various frequencies and depths.
In Fig. \ref{fig5} we show the case of the incoming pulse of the form given by Eq. (23) again with pulse width $\tau$ and the chirp depth $g\sin^2\theta$ but with the frequency equal to $\Delta'=1.2\times10^{-9}$ (while the chirp frequency of the control field was $\Delta=2\times10^{-9}$). One can see that the spectrum structure has evolved along the sample so that the spectrum, initially built of peaks located at multiples od $\Delta'$,  is transformed into one built of peaks located at multiples od $\Delta$. We have checked that the peaks' heights of the outgoing pulse differ from those of the pulse having the modulation frequency equal to $\Delta$ by the factor of 0.225 which is the overlap between the normalized pulses calculated using the above formula with $\tau_1=\tau_2=\tau$, $g_1=g_2=g\sin^2\theta$, $\Delta_1=\Delta'$, $\Delta_2=\Delta$. The behavior of the pulse in space-time is shown in Fig. \ref{fig5b}: again an initially oscillating structure gradually turns into a one-peak structure.

\begin{figure}[H]
\includegraphics[scale=0.6]{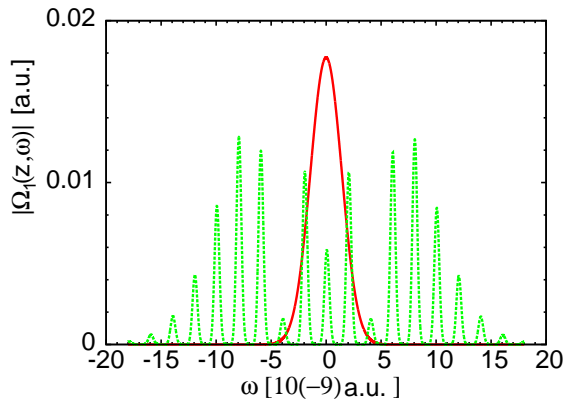}
\caption {\label{fig6}  The pulse spectrum for an incoming Gaussian pulse of width $\tau=10^9$, other parameters as in Fig. \ref{fig2}; the incoming pulse, solid red line (the value has been reduced by the factor of 10; the pulse for $z=10^{11}$, dashed green line.}
\end{figure}

\begin{figure}[H]
\includegraphics[scale=0.6]{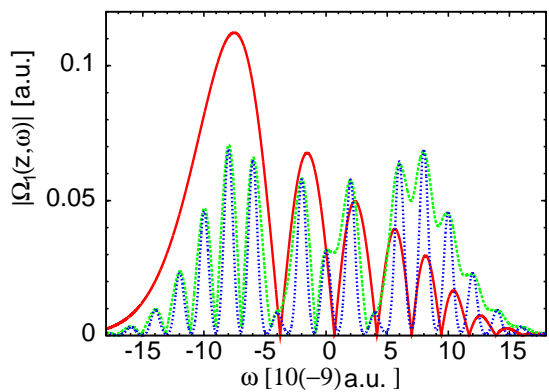}
\caption {\label{fig7}  The pulse spectrum for an incoming chirped pulse of width $\tau=10^9$, chirp depth $g\sin^2\theta$ ($g=5$) and chirp frequency $2\times10^{-9}$, other parameters as in Fig. \ref{fig2}; the incoming pulse, solid red line; the pulse for $z=8\times10^9$, dashed green line; $z=4\times10^{10}$, short-dashed blue line.}
\end{figure}

We next consider the case of a spectrally wide incoming pulse, so that its width is larger than the modulation frequency. In Fig. \ref{fig6} we show the spectrum of the incoming pulse (reduced by the factor of 10) and that of the outgoing pulse for a longer sample ($z=10^{11}$). At the end of the sample we obtain a pulse resembling that of Fig \ref{fig2}, i.e. a sequence of peaks separated by the chirp frequency, of heights proportional to the values of the Bessel functions but significantly narrowed. The width of the peaks is similar to that of a Gaussian pulse propagating in the medium irradiated by an unchirped control field of the Rabi frequency $|\Omega_2|$; in that case the pulse does not completely fit the transparency window corresponding to the unchirped pulse, so its parts of frequencies located at the wings of the spectrum are absorbed and the spectrum is narrowed. The approximate approach based on the assumptions of adiabaticity and first of all a lack of absorption is not allowed any more. The peaks at Fig. \ref{fig6} are additionally shifted; they move to find their final positions at integer values of $\Delta$ only after the pulse has traveled a sufficiently long way.
A similar behavior is even better seen if the incoming pulse is a chirped one. In Fig. \ref{fig7} one can see how the spectrum of the original pulse becomes adjusted: again a significant narrowing of the structures occurs and the peaks' final heights are proportional to the values of the Bessel functions. In space-time the pulse travels as a one-peak structure, significantly absorbed.

For a larger amplitude of the control field, i.e. for a wider transparency window corresponding to an unchirped control pulse, a wide Gaussian probe pulse again adjust the shape of its spectrum to that of chirped control field but because the medium is more transparent and the pulse interacts more weakly, the process of shape adjusting  requires a longer way. We have checked that for the probe pulse of Fig. \ref{fig6} and the control field amplitude $\Omega_2=3\times10^{-8}$ the shape of the probe pulse stabilizes at $z$ being of order of $2\times10^{12}$ (not shown).
\begin{figure}[H]
\includegraphics[scale=0.6]{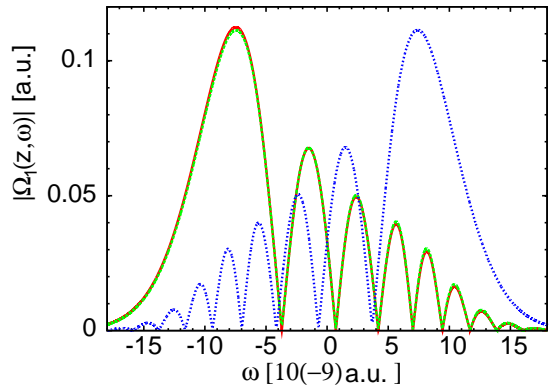}
\caption {\label{fig8}  The pulse spectrum for an incoming chirped pulse as in Fig. \ref{fig6} for a stronger control field $\Omega_2=3\times10^8$, other parameters as in Fig. \ref{fig2}; the incoming pulse at $z=0$ solid red line, the pulse spectrum for $z=4\times 1.53\times10^{9}$ dashed green line (almost coincides with that for $z=0$); for $z=5\times 1.53\times10^9$, short-dashed blue line.}
\end{figure}
In Fig. \ref{fig8} we show the spectrum of a chirped probe pulse in the case of a stronger control field $\Omega_2=3\times10^8$. The spectrum of the incoming pulse is the same  as in Fig. \ref{fig7}. However, now due to a wider transparency window corresponding to an unchirped control field, the pulse propagates almost unabsorbed. Because for a given frequency $\omega$ more than one term in the sum of Eq. (11), as well as Eq. (24), contribute, the spectrum is sensible to the position $z$ inside the sample due to $z$-dependent phase factors. As it follows from Eq. (24), the period of oscillations in $z$ is $2\pi c \cos^2\theta/\Delta$ equal for our data $z_0=3.06 \times 10^9$.
At $z=0$ as well as for even multiples of $z_0/2$ the spectrum is proportional to $\sum_n J_n\exp[-\frac{1}{4}(\omega+n\Delta)^2\tau^2]$. For odd multiples of this value it is proportional to
\begin{eqnarray}
&&\sum_n J_n\exp(i\pi)\exp[-\frac{1}{4}(\omega+n\Delta)^2\tau^2]=\nonumber\\
&&\sum_n J_{-n}\exp[-\frac{1}{4}(\omega+n\Delta)^2\tau^2]=\nonumber\\
&&\sum_n J_n\exp[-\frac{1}{4}(-\omega+n\Delta)^2\tau^2].
\end{eqnarray}
This means that at the position being an odd multiple o $z_0/2$ is a symmetric reflection of the spectrum at $z=0$ as can indeed be seen in Fig. \ref{fig8}. The pulse shape in space-time travels unabsorbed along the sample.

\section{Conclusions}
We have studied a propagation of a weak probe pulse in an atomic medium in the $\Lambda$ configuration with the control field having a constant amplitude and a periodically varying phase, without any restrictions on the value of the modulation frequency. We concentrated mainly on the pulse's spectrum. Numerical results have been compared with those obtained with absorption and nonadiabatic effects being neglected. We have demonstrated that there exist a class of probe pulses for which the medium is transparent: they are built of a phase term with an appropriate chirp (one with the same chirp frequency as the control field and an appropriate depth) and slowly varying envelope. A necessary condition is that the spectrum of the latter fits the transparency window corresponding to the unchirped control field. If this condition is not satisfied due to the pulse having a different chirp, the pulse can be, using a projection method, represented as a sum of that with the appropriate chirp and of the rest; the former part is transmitted while the latter - absorbed.
If the pulse envelope is spectrally too wide, only a part of the pulse is transmitted, so that the resulting spectrum is built of peaks corresponding to the chirp but narrowed as would be the sole pulse envelope in an EIT medium with an insufficiently wide transparency window. In certain conditions the pulse spectra may oscillate along the sample.
\section{acknowledgement}
Bessel functions were computed using the routine due to Shanjie Zhang, Jianming Jin,
Computation of Special Functions, Wiley, 1996, ISBN: 0-471-11963-6, LC: QA351.C45.


\begin{thebibliography}{100}
\bibitem{harris1}
S. E. Harris, J. E. Field, and A. Imamo\~{g}lu, 
Phys. Rev. Lett. {\bf 64}, 1107 (1990).
\bibitem{harris2}
K.-J. Boller, A. Imamo\~{g}lu, and S. E. Harris,
Phys. Rev. Lett. {\bf 66}, 2593-2596 (1991).
\bibitem{EIT}
M. Fleischhauer, A. Imamo\~{g}lu, and J. P. Marangos, 
Rev. Mod. Phys. {\bf 77}, 633-673 (2005).
\bibitem{lukin}
M. D. Lukin, 
Rev. Mod. Phys. {\bf75}, 457-472 (2003).
\bibitem{shore1}
B. Shore, {\em Manipulating Quantum Structures Using Laser Pulses} (Cambridge University Press, 2011).
\bibitem{shore2}
B. Shore, 
Acta Physica Slovaca {\bf 58}, 243-486 (2008).
\bibitem{hau}
C. Liu, Z. Dutton, C. H. Behroozi, and L. V. Hau, 
Nature, {\bf 409}, 490-493 (2001).
\bibitem{phillips}
D. F. Phillips, A. Fleischhauer, A. Mair, R. L. Walsworth, and M. D. Lukin, 
Phys. Rev. Lett. {\bf 86}, 783-786 (2001).
\bibitem{tur}
A.V. Turukhin, V. S. Sudarshanam, M. S. Shahriar, J. A. Musser, B. S. Ham, and P. R. Hemmer,
Phys. Rev. Lett. {\bf 88}, 023602 (2002).
\bibitem{kosachiov1}
D. V. Kosachiov, G. G. Matisov, and Y. U. Rozhdestvensky,
J. Phys. B: At. Mol. Opt. Phys. {\bf 25}, 2473-88 (1992).
\bibitem{cerb1}
E. Cerboneschi, and E. Arimondo, 
Phys. Rev. A {\bf 52}, R1823-6 (1995).
\bibitem{cerb2}
E. Cerboneschi, and E. Arimondo, 
Phys. Rev. A{\bf 54}, 5400-9 (1996).
\bibitem{korsunsky}
A. A. Korsunsky and D. V. Kosachiov, 
Phys. Rev. A {\bf 60}, 4996-5009 (1999).
\bibitem{mahmoudi}
M. Mahmoudi and J. Evers, 
Phys. Rev. A {\bf 74}, 063827 (2006).
\bibitem{eilam1}
A. Eilam, A. D. Wilson-Gordon, and H. Friedmann, 
Opt. Lett. {\bf 34}, 1834-1836 (2009).
\bibitem{scully}
Hebin Li, V. A. Sautenkov, Y. V. Rostovtsev, G. R. Welch, P. R. Hemmer,
and M. O. Scully, 
Phys. Rev. A {\bf 80}, 023820 (2009).
\bibitem{my13}
J. Koroci\'nski, A. Raczy\'nski, J. Zaremba, and S. Zieli\'nska-Kaniasty,
J. Opt. Soc. Am. B {\bf 30}, 1517-1523 (2013).
\bibitem{my14}
A. Raczy\'nski, J. Zaremba, and S. Zieli\'nska-Raczy\'nska,
J. Opt. Soc. Am. B {\bf 31}, 2965-2972 (2014).
\bibitem{salloum}
T. Abi-Salloum, J. P. Davis, C. Lehman, E. Elliott, and F. A. Narducci, 
J. Mod. Opt. {\bf 54}, 2459-2471 (2007).
\bibitem{sautenkov2}
W. A. Sautenkov, H. Li, Y. V. Rostovtsev, G. R. Welch, J. P. Davis, F. A. Narducci, and M. O. Scully, 
J. Mod. Opt. {\bf 56}, 975-979 (2009).
\bibitem{kiffner}
M. Kiffner and T.N. Dey, 
Phys. Rev. A {\bf 79}, 023829 (2009).
\bibitem{my15}
A. Raczy\'nski, J. Zaremba, and S. Zieli\'nska-Raczy\'nska, 
J. Opt. Soc. of Am. B {\bf 32}, 1229-1236 (2015).
\bibitem{fleischhauer2}
M. Fleischhauer and M. D. Lukin, 
Phys. Rev. Let. {\bf 84}, 5094-5097 (2000).
\end{thebibliography}
\end{document}